\documentclass[aps, prl, 10 pt, notitlepage, nofootinbib, twocolumn, letterpaper, superscriptaddress, preprintnumbers, longbibliography, floatfix]{revtex4-1}

\pdfoutput=1
\usepackage{tikz}
\usepackage[compat=1.0.0]{tikz-feynman}

\usepackage[utf8]{inputenc}
\usepackage[normalem]{ulem}
\usepackage{graphicx}
\usepackage{dcolumn}
\usepackage[colorlinks=true,allcolors=purple]{hyperref}
\usepackage{url}
\usepackage{enumerate}
\usepackage{slashed}

\usepackage{slashed,multirow,relsize,soul,feynmp-auto,tikz}
\usepackage{color}
\usepackage{mathrsfs} 
\usepackage{amsmath}
\usepackage{cancel}
\usepackage{bbold}
 \usepackage{mathrsfs}
\usepackage{braket}
\usepackage{physics}
\usepackage{multirow}
\usepackage[capitalize]{cleveref}
\usepackage{xspace}

\usepackage{fontawesome} 
\definecolor{nicegreen}{rgb}{0., 0.75, 0.46}

\definecolor{palatinate}{rgb}{0.494, 0.192, 0.482}
\definecolor{blue-violet}{rgb}{0.33, 0.17, 0.89}

\begin{document}
\preprint{MI-HET-847}
\title{Enabling Strong Neutrino Self-interaction with an Unparticle Mediator}

\author{Saeid Foroughi-Abari}
\email{saeidf@physics.carleton.ca}
\affiliation{Department of Physics, Carleton University, Ottawa, ON K1S 5B6, Canada}
\author{Kevin J. Kelly}
\email{kjkelly@tamu.edu}
\affiliation{Department of Physics and Astronomy, Mitchell Institute for Fundamental Physics and Astronomy, Texas A\&M University, College Station, TX 77843, USA}
\author{Mudit Rai}
\email{muditrai@tamu.edu}%
\affiliation{Department of Physics and Astronomy, Mitchell Institute for Fundamental Physics and Astronomy, Texas A\&M University, College Station, TX 77843, USA}
\author{Yue Zhang}
\email{yzhang@physics.carleton.ca}
\affiliation{Department of Physics, Carleton University, Ottawa, ON K1S 5B6, Canada}
\date{\today}

\begin{abstract}
Recent explorations of the cosmic microwave background and the large-scale structure of the universe have indicated a preference for sizable neutrino self-interactions, much stronger than what the Standard Model offers. When interpreted in the context of simple particle-physics models with a light, neutrinophilic scalar mediator, some of the hints are already in tension with the combination of terrestrial, astrophysical and cosmological constraints. We take a novel approach by considering neutrino self-interactions through a mediator with a smooth, continuous, spectral density function. We consider Georgi's unparticle with a mass gap as a concrete example and point out two useful effects for mitigating two leading constraints. 
1) The Unparticle is ``broadband'' -- it occupies a wide range of masses which allows it to pass the early universe constraint on effective number of extra neutrinos ($\Delta N_{\rm eff.}$) even if the mass gap lies below the MeV scale.
2) Scattering involving unparticles is less resonant -- which lifts the constraint set by IceCube based on a recent measurement of ultra-high-energy cosmogenic neutrinos. Our analysis shows that an unparticle mediator can open up ample parameter space for strong neutrino self-interactions of interest to cosmology and serves a well-motivated target for upcoming experiments.
\end{abstract}

\maketitle

\textbf{Introduction --}
Neutrinos are remnants of the big bang and play an important role in every milestone of the early universe, from primordial nucleosynthesis (BBN) to the formation of the cosmic microwave background (CMB) and large scale structures.
The corresponding cosmological observations provide a powerful lens into the fundamental properties of neutrinos.
Novel neutrino self-interactions are such a target.
Recently, there has been tantalizing evidence of frequent self-scattering among neutrinos and delayed free-streaming until close to matter-radiation equality, suggested by explorations of the CMB~\cite{Kreisch:2019yzn,Das:2020xke,RoyChoudhury:2020dmd,Brinckmann:2020bcn,Kreisch:2022zxp,Das:2023npl} and the matter power spectrum~\cite{He:2023oke,Camarena:2023cku,Camarena:2024zck,Pal:2024yom, Racco:2024lbu}. The favored neutrino self-interaction is characterized by a sizable ``Fermi constant,'' $G_{\rm eff.} \sim (10-100\,{\rm MeV})^{-2}$, much larger than that of the weak interaction.
Over the next decade, a new generation of cosmological experiments will deliver copious data, enabling unprecedented precision and opportunities to pin down the new physics behind self-interacting neutrinos~\cite{Berryman:2022hds,Gerbino:2022nvz,Chou:2022luk}.

The simplest and most vetted model for neutrino self-interactions introduces a light scalar particle that couples exclusively to neutrinos~\cite{Berryman:2018ogk, Blinov:2019gcj, Berryman:2022hds}.
There are various complementary probes and constraints on the scalar mediator and in turn the neutrino self-interaction parameter space. 
For electron and muon neutrinos, couplings to a neutrinophilic scalar are tightly constrained by exotic charged-meson decay measurements if the scalar has a mass below 400 MeV~\cite{Barger:1981vd, Brdar:2020nbj, Lyu:2020lps, Dev:2024twk}. In contrast, the coupling between the scalar mediator and tau neutrinos is a blind spot of searches and is allowed to be much stronger. 
A flavor-independent constraint arises from the measurement of effective number of neutrino species using BBN, CMB and matter power spectrum~\cite{Pitrou:2018cgg,DESI:2024hhd, Escudero:2024uea}, $\Delta N_{\rm eff.}\lesssim 0.3$. It translates into a lower bound on the mediator mass, 2.0 (3.1) MeV for a real (complex) scalar particle, assuming it equilibrates with neutrinos before the weak interaction decoupling.
Interestingly, recent analyses of the CMB datasets show preference for flavor-specific strong self-interactions of neutrinos~\cite{Das:2020xke,Das:2023npl}.

Exploration of ultra high-energy (UHE) cosmic neutrinos with the IceCube Neutrino Observatory provides another powerful test of neutrino self-interactions. Collisions of UHE neutrino with the cosmic neutrino background (C$\nu$B), through the same fundamental process as in the early universe, can distort the observed UHE neutrino energy spectrum~\cite{Ioka:2014kca,Ng:2014pca,Ibe:2014pja,Kamada:2015era}. A recent analysis~\cite{Esteban:2021tub} of 7.5-year IceCube data~\cite{IceCube:2020wum} constraints $\nu_\tau$-philic scalars with mass below 40 MeV and substantially narrows down the room for sizable neutrino self-interactions. 

The above status motivates us to ask the question: can the strong self-interacting neutrino scenario be accommodated in any fundamental theories with the IceCube and $\Delta N_{\rm eff.}$ constraints lifted, so that they will continue to guide upcoming cosmological explorations?

\begin{figure*}[!htbp]
    \begin{center}
        \includegraphics[width=0.8\linewidth]{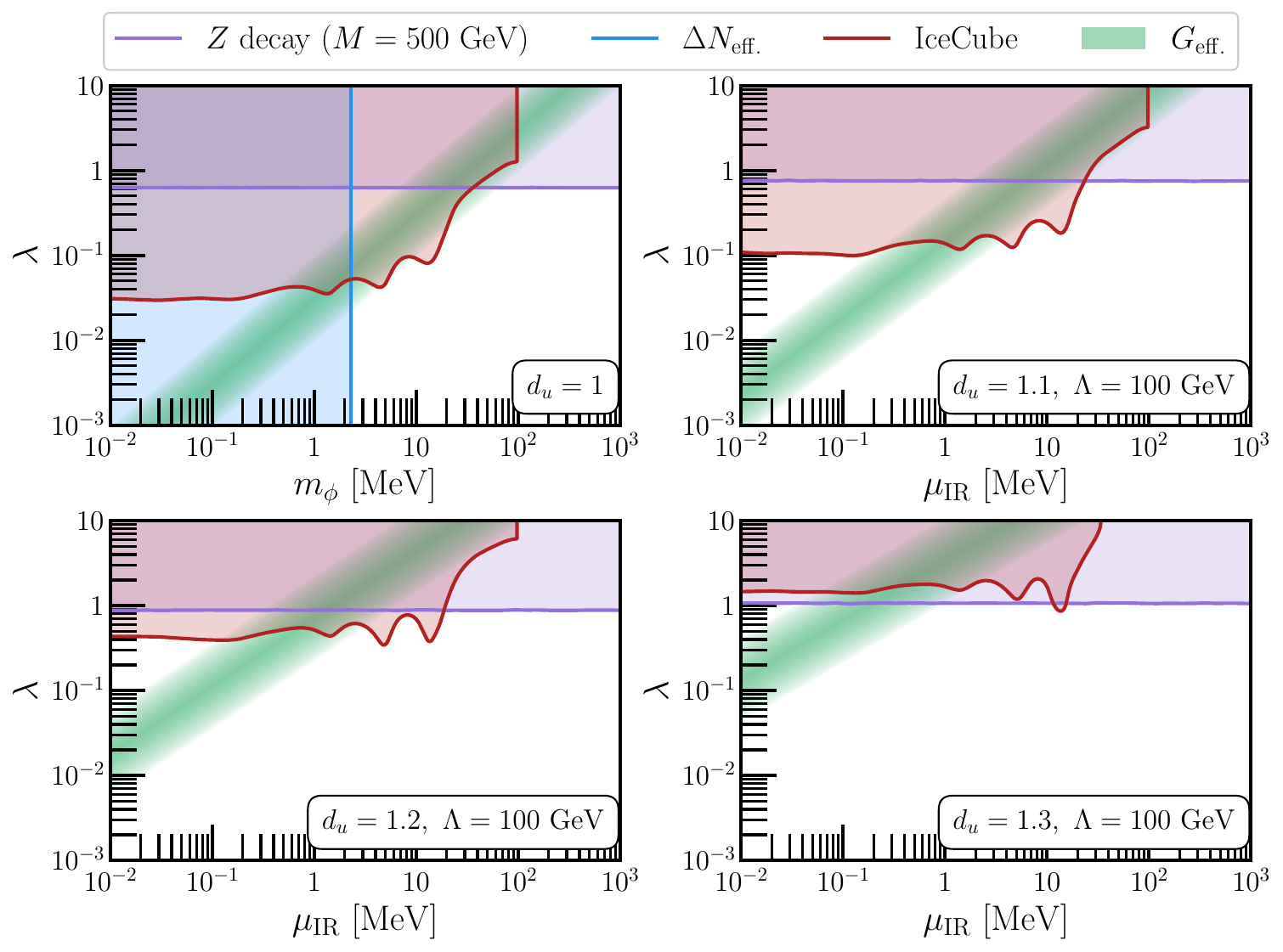}
        \caption{Self-interacting neutrino parameter space for particle (top-left) and unparticle (others) mediator cases, with $d_u=1, 1.1, 1.2, 1.3$. The green shaded regions correspond to $G_{\rm eff.} \in [(10-100) \,{\rm MeV}]^{-2}$ and delayed onset of neutrino free streaming, which is a target for upcoming cosmological explorations. The red, blue, and purple-shaded regions are excluded by IceCube, $\Delta N_{\rm eff.}$, and invisible $Z$-width constraints, respectively. We fix parameters $\Lambda=100\,$GeV and $M=500\,$TeV. Our findings suggest that an unparticle mediator can open up ample parameter space to accommodate strong neutrino self-interactions.}\label{fig:All_du}
    \end{center}
\end{figure*}

\textbf{Gapped unparticle mediator -- }
In this Letter, we take a new approach by considering neutrino self-interaction through the exchange of a mediator with a continuous spectral density function. As a simple and calculable model, we will consider Georgi's unparticle~\cite{Georgi:2007ek,Georgi:2007si} augmented by a mass gap, which is used to describe a conformal-invariant hidden sector within a certain energy window. 
More generally, continuum excitations can arise from extra dimensional theories or hidden gauge sectors~\cite{Arkani-Hamed:1998wff,Cacciapaglia:2008ns,Falkowski:2008fz,Cabrer:2009we,Chacko:2020zze,Csaki:2021gfm}.

To mediate neutrino self-interactions, we write down the Yukawa-interacting Lagrangian between an unparticle field $\mathcal{U}$ and the Standard Model tau neutrino ($\nu_\tau$),
\begin{align}\label{eq:Lagrangian}
    \mathcal{L}_{\rm int} = \frac{\lambda}{2\Lambda^{d_u-1}} \bar{\nu}^c_\tau \mathbb{P}_L \nu_\tau \mathcal{U} + \mathrm{h.c.}\ ,
\end{align}
where $\nu^c_\tau$ is the right-handed anti-neutrino and charged under $SU(2)_L$.
$\mathcal{U}$ can be a real or complex field with zero spin. Its mass dimension is bounded from below, $d_u>1$ by unitarity~\cite{Qualls:2015qjb}.

We consider a gapped unparticle with K{\"a}ll{\'e}n–Lehmann spectral density function~\cite{Fox:2007sy,Cacciapaglia:2007jq}
\begin{equation}\label{eq:KL}
    \rho_{\rm KL}(m^2) = A_{d_u } \left(m^2 - \mu_{\rm IR}^2\right)^{d_u - 2} \Theta\left(m^2- \mu_{\rm IR}^2\right) \ ,
\end{equation}
 where $\Theta$ is the unit step function, and $\mu_{\rm IR}$ is the mass gap of the hidden-sector continuum states. 
The unparticle description is valid up to the ultraviolet cutoff scale $\mu_{\rm UV}$ of the hidden-sector conformal window. We will ensure that it remains above all physical energy scales considered in this work.
The coefficient $A_{d_u}$ takes the form~\cite{Georgi:2007ek}
\begin{equation}\label{eq:Adu}
A_{d_u} = \frac{16\pi^{5/2}}{\left(2\pi\right)^{2d_u}}
\frac{\Gamma(d_u + \frac{1}{2})}{\Gamma\left(d_u - 1\right) \Gamma\left(2d_u\right)}  \ ,
\end{equation}
where $\Gamma$ is the Euler gamma function.
The particle limit corresponds to $d_u\to 1$, where the spectral density reduces to a $\delta$-function, $\rho_{\rm KL}(m^2)\to (2\pi)\delta(m^2-\mu_{\rm IR}^2)$.

The presence of the interaction term~\cref{eq:Lagrangian} introduces a redundancy among the unparticle parameters.
One has the option of redefining the field $\mathcal{U}$ to absorb the factor $\Lambda^{1-d_u}$. In that case, $\mathcal{U}$ has dimension 1 but still describes an unparticle, since the redefinition only rescales the spectral density function in~\cref{eq:KL} by a factor of $1/\Lambda^{2(d_u-1)}$. Requiring the spectral function to always normalize to unity, $\int_{\mu_{\rm IR}}^{\mu_{\rm UV}} \rho_{\rm KL}(m^2)dm^2/(2\pi)= \Lambda^{2(d_u-1)}$, leads to
\begin{equation}\label{eq:redundancy}
\mu_{\rm UV}^2 - \mu_{\rm IR}^2 = 16\pi^2  \Gamma(d_u)^{\frac{2}{d_u-1}} \Lambda^2 \ .
\end{equation}
This normalization condition spreads one particle's worth spectral density into a continuous mass distribution between $\mu_{\rm IR}$ and $\mu_{\rm UV}$.
It allows for an apples-to-apples comparison between the particle and unparticle mediators. Independent parameters for gapped unparticle mediating neutrino self-interactions are
\begin{align*}
\{\lambda, \Lambda, \mu_{\rm IR}, d_u\}.
\end{align*}

\textbf{Self-interacting Neutrinos for Cosmology --}
The Feynman propagator for the unparticle reads~\cite{Rajaraman:2008bc,Delgado:2008gj}
\begin{align}\label{eq:propagator}
\begin{split}
    D_F(q) 
    &= \int_{\mu_{\rm IR}^2}^{\mu_{\rm UV}^2} \frac{dm^2}{2\pi} \rho_{\rm KL}(m^2) \frac{i}{q^2 - m^2 + i\varepsilon} \\
    &\simeq \frac{i B_{d_u}}{(q^2 - \mu_{\rm IR}^2)^{2-d_u} - i {\rm Im}\Sigma(q^2)} \ ,
\end{split}    
\end{align}
where $B_{d_u} = (-1)^{d_u-2} A_{d_u}/[2\sin(\pi d_u)]$. The second line is acquired by assuming $\mu_{\rm UV}^2\gg\mu_{\rm IR}^2, |q^2|$, and ${\rm Im}\Sigma(q^2) = -{|\lambda|^2 B_{d_u}q^2}/({16\pi \Lambda^{2d_u-2}})$ resums the absorptive part of neutrino bubble contributions to the unparticle's self-energy from the interaction in~\cref{eq:Lagrangian}.

Neutrino self-interactions relevant for late-time cosmological observations such as CMB and structure formation occur at very low center-of-mass energies, $|q^2| \ll \mu_{\rm IR}^2$.  Integrating out the unparticle field, the neutrino self-interaction strength derives from 
\begin{align}
\mathcal{L}_{\rm eff.} = G_{\rm eff.}(\bar\nu_\tau \gamma^\mu \mathbb{P}_L \nu_\tau) (\bar\nu_\tau \gamma_\mu \mathbb{P}_L \nu_\tau) \ , 
\end{align}
with the dominant contribution coming from unparticle states close to the mass gap,
\begin{align}\label{eq:Geff}
    G_{\rm eff.} = \frac{|B_{d_u}\lambda^2|}{\mu_{\rm IR}^2}\left(\frac{\mu_{\rm IR}}{\Lambda}\right)^{2d_u-2} \ .
\end{align}
The rough parameter space of interest to cosmology, $G_{\rm eff.} \sim \left(10-100~\mathrm{MeV}\right)^{-2}$, is shown by the green region in~\cref{fig:All_du}.
In the $d_u\to1$ limit, $B_{d_u}\to1$ and $G_{\rm eff.}\to |\lambda|^2/\mu_{\rm IR}^2$ which corresponds to the exchange of a scalar particle with mass $\mu_{\rm IR}$~\cite{Blinov:2019gcj}.

\textbf{Lifting $\Delta N_{\rm eff.}$ Constraint --}
The unparticle-neutrino interaction can bring the unparticle states into thermal equilibrium in the early universe, leading to a contribution to $\Delta N_{\rm eff.}$. 
At temperature $T$, the energy density of a fully thermalized unparticle is
\begin{align}
\varrho_u (T) = \frac{\int_{\mu_{\rm IR}^2}^{\mu_{\rm UV}^2}\frac{dm^2}{2\pi} \rho_{\rm KL}(m^2) g\int\frac{d^3 \vec{p}}{(2\pi)^3} \frac{E_{\vec{p}}}{\exp\left( E_{\vec{p}}/T\right)-1} }{\int_{\mu_{\rm IR}^2}^{\mu_{\rm UV}^2} \frac{dm^2}{2\pi} \rho_{\rm KL}(m^2)} \ ,
\label{eq: rho}
\end{align}
where $E_{\vec{p}} = \sqrt{|\vec{p}|^2+m^2}$, and the factor $g=1~(2)$ if $\mathcal{U}$ is a real (complex) field. 
The contribution to $\Delta N_{\rm eff.}$ during BBN time be estimated using $\Delta N_{\rm eff.} = \varrho_u (T_{\rm BBN}) \left({7\pi^2 T_{\rm BBN}^4}/{120}\right)^{-1}$,
where $T_{\rm BBN}\simeq 0.8$\,MeV is the weak-interaction decoupling temperature. 

\begin{figure}[t]
    \centering
    \includegraphics[width=0.75\linewidth]{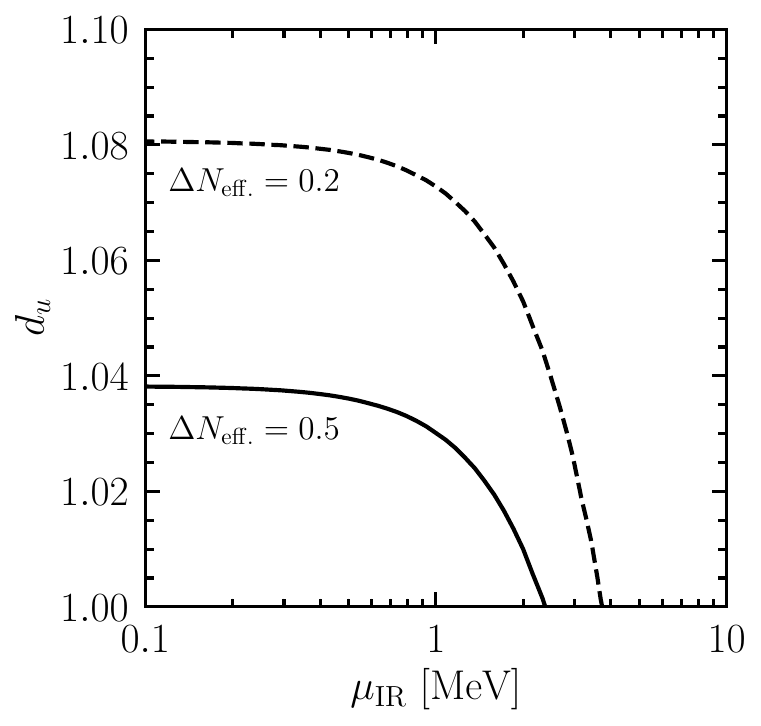}
    \caption{Contribution to $\Delta N_{\rm eff.}$ during the epoch of BBN from the unparticle.  
    The solid (dashed) line corresponds to $\Delta N_{\rm eff.}=0.5 \,(0.2)$ for comparison. Here we set $\Lambda = 100$~GeV.\label{fig:Neff}}
\end{figure}

Much of the parameter space of interest for strong self-interactions satisfies $\mu_{\rm IR} \ll T_{\rm BBN} \ll \mu_{\rm UV}$, so we find it instructive to determine the unparticle contribution to $\Delta N_{\rm eff.}$ analytically in this regime:
\begin{equation}\label{eq:Neff}
\begin{split}
\Delta N_{\rm eff.} &\simeq  \frac{1920 d_u(2d_u+1)\zeta(2d_u+2)}{7 (2\pi)^{2d_u+2}}\cdot g\left(\frac{T_{\rm BBN}}{\Lambda}\right)^{2d_u-2} \ ,  
\end{split}
\end{equation}
where $\zeta$ is the Riemann zeta function. We provide more details for this derivation in Appendix~\hyperref[appendix:Neff]{A} (see also~\cite{Chen:2007qc}).
Taking $d_u\to1$ limit of the above result corresponds to a massless scalar particle contribution to $\Delta N_{\rm eff.} = 4g/7$. 
In contrast, for $d_u>1$, $\Delta N_{\rm eff.}$ contains an extra suppression factor $\left({T_{\rm BBN}}/{\Lambda}\right)^{2d_u-2}$ compared to the particle case.
The origin of this factor can be understood, as the unparticle has a continuous mass distribution between $\mu_{\rm IR}$ and $\mu_{\rm UV}$. At temperature $T$, only the states within the window $\mu_{\rm IR}<m\lesssim T$ are populated and contribute to $\Delta N_{\rm eff.}$. The $T\ll m<\mu_{\rm UV}$ part, although carrying a large weight in the spectral function, is Boltzmann-suppressed. 
If the unparticle emergence scale $\mu_{\rm UV} \sim \Lambda$ is much higher than $T_{\rm BBN}$, 
the suppression factor allows all current constraints on $\Delta N_{\rm eff.}$ from BBN and CMB to be lifted with $d_u>1.1$. This behavior is shown in~\cref{fig:Neff}, in which we perform the complete calculation for $\Delta N_{\rm eff.}$ as a function of $\mu_{\rm IR}$ and $d_u$, fixing $\Lambda = 100$~GeV.

It is useful to comment on the parametrical interplay between $G_{\rm eff.}$ and $\Delta N_{\rm eff.}$ derived in Eqs.~\eqref{eq:Geff} and \eqref{eq:Neff}, respectively. They contain a common factor $1/\Lambda^{2d_u-2}$ which seems to imply that making $\Delta N_{\rm eff.}$ smaller comes at the cost of suppressing the neutrino self-interaction. However, it is crucial to note that the dependence on a small $\mu_{\rm IR}$ drops out of $\Delta N_{\rm eff.}$ but not $G_{\rm eff.}$.
An unparticle mediator with a mass gap below the MeV scale can open up parameter space to accommodate strong neutrino self-interactions of interest to late-time cosmology. This interplay is demonstrated in~\cref{fig:All_du}.
The point that BBN allows for a sub-MeV mass gap holds generically for dark sectors made of unparticle states, extending beyond neutrinophilic models.

\textbf{Lifting IceCube Constraint --}
The astrophysical frontier offers another useful probe of the neutrino self-interaction, using the scattering between a PeV-scale cosmogenic neutrino and one from the C$\nu$B.
For neutrino masses around 0.1 eV, the center-of-mass energy of the scattering lies around 10 MeV, allowing for on-shell production of the neutrinophilic mediator. The subsequent decay of the mediator produces less energetic neutrinos, introducing novel features such as dips and bumps in the UHE cosmic neutrino spectrum~\cite{Esteban:2021tub}.

\begin{figure}[t]
    \centering
    \includegraphics[width=1\linewidth]{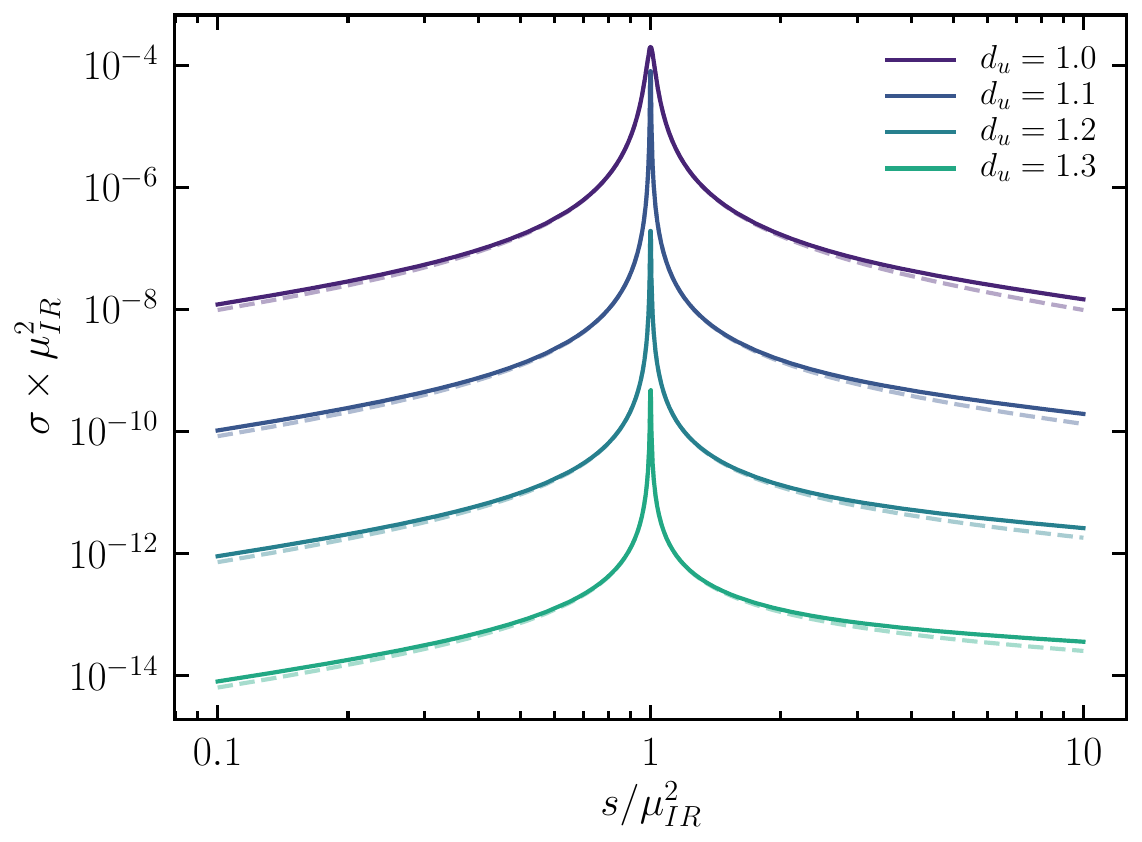}
    \caption{Near-resonance behavior of the neutrino self-interacting cross section for particle and unparticle mediator cases. The solid (dashed) curve corresponds to the total ($s$-channel only) cross section.}\label{fig:propagator}
\end{figure}

In the model \eqref{eq:Lagrangian}, $\nu\nu\to\nu\nu$ scattering can occur via the mediator exchange in the $s$-channel, whereas $\nu\bar\nu\to\nu\bar\nu$ occurs in the $t$-channel. For neutrinos traveling across cosmological distances, the quantum coherence for flavor oscillations is already lost and one must work in the neutrino mass basis for the initial and final states.
The differential cross sections are
\begin{eqnarray}\label{eq:UHECnuB}
&&\hspace{-0.5cm}\frac{d\sigma_{ij\to kl}}{d t} = |U_{\tau i}|^2 |U_{\tau j}|^2 |U_{\tau k}|^2 |U_{\tau l}|^2  \frac{|\lambda|^4 |B_{d_u}|^2}{16\pi \Lambda^{4(d_u-1)}} \\
&&\hspace{-0.3cm}
\times \left[ \frac{1}{\left( s-\mu_{\rm IR}^2 \right)^{4-2d_u} + |{\rm Im}\Sigma(s)|^2} + \frac{t^2/s^2}{\left( t -\mu_{\rm IR}^2 \right)^{4-2d_u}} \right] \ ,\nonumber
\end{eqnarray} 
where $s, t$ are the Mandelstam parameters, $U$ is the leptonic mixing matrix and $i,j,k,l=1,2,3$ are used to label neutrino or anti-neutrino mass eigenstates.
In the particle limit ($d_u=1$), the differential cross section features a Breit-Wigner resonance at $s = \mu_{\rm IR}^2$. In contrast, the resonance effect shrinks significantly for the unparticle case ($d_u>1$), as shown in Fig.~\ref{fig:propagator}. As a result, the reaction cross section between the UHE and C$\nu$B neutrinos is reduced and the corresponding IceCube constraint on the neutrino self-interaction is weakened.

\begin{figure}[t]
    \centering
    \includegraphics[width=1\linewidth]{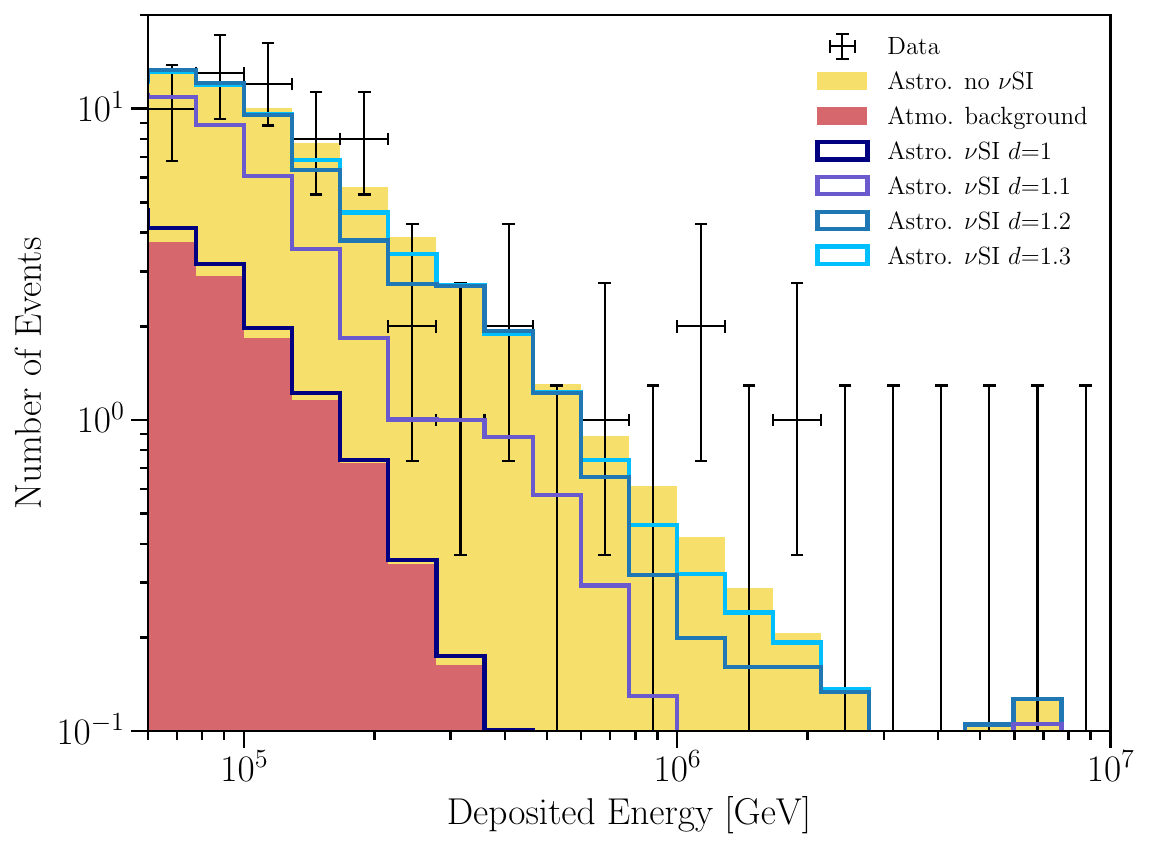}
    \caption{IceCube deposited energy distributions: HESE 7.5-year data is shown as crosses, with best-fit expectations (no new physics) represented as stacked histograms for astrophysical neutrinos (gold) and atmospheric background (red). A stronger $\nu$SI model with a gapped unparticle mediator and sufficiently large $d_u$ revives scenarios excluded in the particle mediator case (navy). See text for details.
    \label{fig:histogram}}
\end{figure}

To deliver this point home, we solve the UHE neutrino transport equation~\cite{Lee:1996fp, Bhattacharjee:1999mup, Ng:2014pca}  taking into account neutrino self-interactions between the source and Earth. We generalize the {\tt nuSIProp} code provided in Ref.~\cite{Esteban:2021tub} to the unparticle mediator case (see appendix~\hyperref[appendix:nusiprop]{B} for details).
The obtained neutrino flux that reaches the Earth is input for the code accompanying the IceCube 7.5 year data release~\cite{HESEData2021} for simulating the number of high-energy starting-events (HESE) in energy bins above 60 TeV. FIG.~\ref{fig:histogram} shows a useful example to compare different self-interacting neutrino scenarios, where we set $\lambda = 0.2$, $\mu_{\rm IR} = 5\,{\rm MeV}$, and $\Lambda = 100\,{\rm GeV}$. 
The UHE astrophysical neutrino flux (source term of the transport equation~\cite{Esteban:2021tub}) is assumed to follow a simple power law, with the spectral index $\gamma_{\rm astro}=2.87$ and normalization factor $\Phi_{\rm astro}=6.37$ held fixed.
The sum of neutrino masses is also fixed to be 0.07\,eV (for demonstrative purposes only; we vary these parameters in our analyses to follow).
The yellow histogram corresponds to the absence of neutrino self-interactions.
The dark-blue dashed curve gives the HESE event prediction in the particle mediator case, whereas the unparticle mediator cases are the purple, cobalt, and light-blue solid curves for $d_u=1.1,1.2,1.3$, respectively.  
Clearly, the particle mediator case exhibits strong distortions to the UHE neutrino flux and is incompatible with data, whereas the effect is much less significant for unparticle mediator with higher $d_u$.

We further take this approach to scan over the $\lambda$ versus $\mu_{\rm IR}$ parameter space for particle and unparticle mediators. For every point, we allow $\left\lbrace\gamma_{\rm astro}, \Phi_{\rm astro}, \sum m_\nu\right\rbrace$ to vary and adopt an MCMC algorithm based on the \texttt{emcee} package~\cite{Foreman_Mackey_2013} to find the best fit to the IceCube data. The $2\sigma$ exclusions are given by the red shaded regions in~\cref{fig:All_du}. 
More details are provided in appendix~\hyperref[appendix:triangleplot]{C}.
Moving from particle to unparticle mediator cases, both the IceCube and BBN constraints are weakened.
Although the self-interacting neutrino region (in green) also moves upward, the constraints retreat faster. As a result, ample parameter space opens up that can accommodate strong neutrino self-interactions.

\textbf{Terrestrial Constraint --}
For completeness, we include a flavor-independent constraint on a neutrinophilic unparticle from the invisible $Z$-boson decay width~\cite{ParticleDataGroup:2024cfk}. With the interacting Lagrangian \eqref{eq:Lagrangian}, $\mathcal{U}$ can contribute to the two-body decay $Z\to\nu_\tau\bar\nu_\tau$ at loop level, and the three-body decays $Z\to\nu_\tau\nu_\tau \mathcal{U}$ and $\bar\nu_\tau\bar\nu_\tau \mathcal{U}^*$,
\begin{widetext}

\begin{align}
\label{eq:7}
    &\delta \Gamma_{Z\to\nu_\tau\bar\nu_\tau} = - \frac{\sqrt{2} G_F M_Z^3 |\lambda|^2}{192\pi^3} \int_{\mu_{\rm IR}^2}^{\mu_{\rm UV}^2} \frac{dm^2}{2\pi} \frac{\rho_{\rm KL}(m^2)}{\Lambda^{2d_u-2}}\int_{0}^{1} dx \int_{0}^{1-x} dy \nonumber\\
&\hspace{2.7cm}\times\mathrm{Re}\left\lbrace \ln\frac{M^2}{m^2} + \ln\left[\frac{M^2}{(1-x-y)m^2 - xyM_Z^2}\right] + \frac{xyM_Z^2}{(1-x-y)m^2 - xyM_Z^2} - \frac{3}{2}  \right\rbrace \ ,\\
\label{eq:8}
    &\Gamma_{Z\to\nu_\tau\nu_\tau \mathcal{U}} + \Gamma_{Z\to\bar\nu_\tau\bar\nu_\tau \mathcal{U}^*} = \frac{\sqrt{2} G_F M_Z^3 |\lambda|^2}{192\pi^3} \int_{\mu_{\rm IR}^2}^{{\rm min}{(\mu_{\rm UV}^2}, M_Z^2)} \frac{dm^2}{2\pi} \frac{\rho_{\rm KL}(m^2)}{\Lambda^{2d_u-2}} \left\lbrace \left[\frac{1}{2}\ln{\left(\frac{M_Z^2}{m^2}\right)} - \frac{11}{6}\right] + \frac{4m^2}{M_Z^2}\left[\ln{\left(\frac{M_Z^2}{m^2}\right)} - \frac{5}{8}\right] \right. \nonumber\\
&\hspace{2.7cm}\left. + \frac{m^4}{M_Z^4} \left[\ln{\left(\frac{M_Z^2}{m^2}\right)} \left(
\ln{\left(\frac{M_Z m}{M_Z^2+m^2}\right)} + \frac{3}{2}\right) + \frac{9}{2}
+{\rm Li}_2\left( \frac{M_Z^2}{M_Z^2+m^2}\right)
-{\rm Li}_2\left( \frac{m^2}{M_Z^2+m^2}
\right)
\right] - \frac{m^6}{6M_Z^6}\right\rbrace \ .
\end{align}
\end{widetext}
These results are obtained with a simple generalization of the decay width results for the particle mediator case.
See appendix~\hyperref[appendix:ZDecay2]{D} for more details.
In Fig.~\ref{fig:All_du}, the invisible $Z$-width measurement bounds the coupling $\lambda$ from above and excludes the purple shaded region.

The cosmologically interesting region in Fig.~\ref{fig:All_du} extends to low $\mu_{\rm IR}$ with very small $\lambda$. In that region of parameter space, the severe constraints from pion and kaon decays that apply for $\nu_e$ and $\nu_\mu$~\cite{Berryman:2022hds} can also be evaded, enabling all neutrinos to self-interact strongly.

\textbf{Outlook --}
The self-interacting neutrino scenario is a new physics target of great interest to cosmological explorations. In this work, we consider a continuum mediator for the self-interaction and show that two leading constraints from $\Delta N_{\rm eff.}$ and IceCube can be lifted 
Although the calculations are done in a phenomenological model of gapped unparticle, 
we expect qualitatively similar results to hold in theories where the mediator is described by a continuous and smooth spectral density function.
Our work broadens the theoretical framework behind self-interacting neutrinos and opens up large parameter space for further tests.

With upcoming experiments, the unparticle (continuum) mediator for strong neutrino self-interaction can be scrutinized in at least three ways. 1) The UHE neutrino observation made by IceCube-Gen2 with a higher exposure will be sensitive to smaller $\lambda$ coupling~\cite{IceCube-Gen2:2020qha,Esteban:2021tub};
2) The future FCC-ee collider can measure the invisible $Z$ width by 8 times more precisely~\cite{FCC:2018byv, FCC:2018evy} using the radiative return channel;
3) With a mass gap below MeV, part of the unparticle states freeze out by decaying into neutrinos at temperatures between BBN and CMB. This will lead to a slightly higher $\Delta N_{\rm eff.}$ for CMB~\cite{Chacko:2003dt, EscuderoAbenza:2020cmq, Kelly:2020aks} and be tested by the upcoming CMB-S4 experiment.

\textbf{Acknowledgement} -- We thank Nikita Blinov, Vera Gluscevic, Nahee Park, Ryan Plestid, Daniel Stolarski and Flip Tanedo for useful discussions. SF and YZ are supported by a Subatomic Physics Discovery Grant (individual) from the Natural Sciences and Engineering Research Council of Canada. KJK and MR are supported by DOE Grant No. DE-SC0010813.
This research was enabled in part by support provided by the Digital Research Alliance of Canada (\url{https://alliancecan.ca}).

\bibliographystyle{apsrev4-1}
\bibliography{main}{}

\clearpage

\onecolumngrid
\section{Supplemental Material}

\subsection{A. $\Delta N_{\rm eff.}$ contribution from unparticle with a small mass gap}\label{appendix:Neff}

In this appendix, we derive an analytic formula for the unparticle contribution to $\Delta N_{\rm eff.}$ in the limit $\mu_{\rm IR} \ll T_{\rm BBN} \ll \mu_{\rm UV}$. While this range does not cover the entire region of interest for unparticles and neutrino self-interactions, we find it demonstrative and can derive a simple, analytic result. When considering a real or complex scalar particle's contribution to $\Delta N_{\rm eff.}$, bringing the particle's mass below $T_{\rm BBN}$ is highly restrictive; instead, when considering an unparticle, we see that even a sub-MeV mass gap $\mu_{\rm IR}$ is allowed once $d_u \neq 1$.

In the above limit, the numerator Eq.~\eqref{eq: rho} in the main text is approximately
\begin{align}
{\rm numerator} = \frac{g A_{d_u}}{2\pi^3}\int_0^\infty m^{2d_u-3} dm \int_0^\infty p^2dp \frac{\sqrt{p^2+m^2}}{\exp\left( \sqrt{p^2+m^2}/T\right)-1} \ ,
\end{align}
where $p$ is the magnitude of the three-momentum vector $\vec{p}$. It is useful to transform to polar coordinates and define $p=R\cos\theta$, $m=R\sin\theta$, and rewrite the above integrals as
\begin{equation}
\begin{split}
{\rm numerator} &= \frac{g A_{d_u}}{2\pi^3}\int_0^{\pi/2} d\theta (\sin\theta)^{2d_u-3} \cos^2\theta \int_0^\infty RdR   \frac{R^{2d_u}}{\exp\left(R/T\right)-1} \\
&=\frac{g A_{d_u} \Gamma(d_u-1)\Gamma(2d_u+2) \zeta(2d_u+2) T^{2d_u+2}}{8\pi^{5/2}\Gamma(d_u+1/2)} \\
&=\frac{4d_u(2d_u+1)\zeta(2d_u+2)}{(2\pi)^{2d_u}} gT^{2d_u+2} \ ,
\end{split}
\end{equation}
where in the last step, we plugged in Eq.~\eqref{eq:Adu}. Using the relation Eq.~\eqref{eq:redundancy} derived for the unparticle field normalization, the denominator of Eq.~\eqref{eq: rho} is simply given by
\begin{align}
\Lambda^{2d_u-2} \ .
\end{align}
As a result,
\begin{equation}
\begin{split}
\Delta N_{\rm eff.} &\simeq  \frac{1920 d_u(2d_u+1)\zeta(2d_u+2)}{7 (2\pi)^{2d_u+2}}\cdot g\left(\frac{T_{\rm BBN}}{\Lambda}\right)^{2d_u-2} \ ,  
\end{split}
\end{equation}
which corresponds to Eq.~\eqref{eq:Neff} in the main text.

\subsection{B. UHE neutrino flux calculation for the unparticle mediator case}\label{appendix:nusiprop}

To derive the UHE astrophysical neutrino flux in the presence of neutrino self-interaction, we use the {\tt nuSIprop} code~\cite{Esteban:2021tub}. In particular, the code evolves the neutrino spectrum by discretizing the neutrino energy and the redshift, as described in their note~\cite{ivan2024nuSIprop}. In this appendix, we present the relevant quantities defined in the note by generalizing the calculation to the unparticle mediator case. 

Using the differential cross section Eq.~\eqref{eq:UHECnuB} in the main text, we can derive the total scattering cross section between the UHE neutrino and C$\nu$B,
\begin{align}\label{Eq:CS}
     \sigma_{ij}(s) &= \frac{ |\lambda|^4 B_{d_u}^2}{16 \pi \Lambda^{4(d_u-1)}} |U_{\tau i}|^2 |U_{\tau j}|^2 \sum_{k,l} |U_{\tau k}|^2 |U_{\tau l}|^2  \\
     & \times\Bigg[ 
      \frac{s}{\left(s - \mu_{\rm IR}^2\right)^{2(2-d_u)} + |{\rm Im}\Sigma(s)|^2} + \frac{ \left(s + \mu_{\rm IR}^2\right)^{2d_u{-}3} \left[(d_u{-}1)(2d_u{-}3)s^2 - (2 d_u{-}3) s\mu_{\rm IR}^2 + \mu_{\rm IR}^4\rule{0mm}{4mm}\right] -\mu_{\rm IR}^{2(2d_u{-}1)}}
     {(d_u{-}1)(2d_u{-}3)(2d_u{-}1)s^2 }
     \Bigg]  \ ,\nonumber
\end{align}
where unitarity of the leptonic mixing matrix implies the sum $\sum_k|U_{\tau k}|^2 =\sum_l |U_{\tau l}|^2 =1$. 
Inside the square bracket, the first term corresponds to $s$-channel $\nu\nu\to\nu\nu$ scattering whereas the second term is for $t$-channel $\nu\bar\nu\to\nu\bar\nu$ scattering. They do not interfere with each other in the $m_\nu^2\ll s$ limit.

The coefficient functions used in the neutrino transport equation are defined as~\cite{ivan2024nuSIprop}
\begin{align}
& \Gamma_i^n \equiv \sum_j \int_{E_{n-1 / 2}}^{E_{n+1 / 2}} \sigma_{i j}\left(E_\nu\right) \mathrm{d} E_\nu =  \sum_j \int_{s^-}^{s^+} \frac{\mathrm{d}s}{2m_j}\sigma_{i j}(s) \ , \\
&
\tilde{\alpha}_{i j}^n \equiv \sum_{j_1, j_2} \int_{E_{n-1 / 2}}^{E_{n+1 / 2}} \mathrm{~d} E_\nu \int_E^{E_{n+1 / 2}} \mathrm{~d} \tilde{E}_\nu \frac{\mathrm{d} \sigma_{j j_1 \rightarrow i j_2}}{\mathrm{~d} E_\nu}\left(\tilde{E}_\nu, E_\nu\right) =  \sum_{j_1, j_2} \int_{t^-}^{t^+}\mathrm{d}t \int_{-t}^{-t^+}\frac{\mathrm{d}s}{2m_{j_1}} \frac{\mathrm{d} \sigma_{j j_1 \rightarrow i j_2}}{\mathrm{~d} t}\left(s, t\right) \ , \\
& \alpha_{i j}^{n, n^{\prime}} \equiv \sum_{j_1, j_2} \int_{E_{n-1 / 2}}^{E_{n+1 / 2}} \mathrm{~d} E_\nu \int_{E_{n^{\prime}-1 / 2}}^{E_{n^{\prime}+1 / 2}} \mathrm{~d} \tilde{E}_\nu \frac{\mathrm{d} \sigma_{j j_1 \rightarrow i j_2}}{\mathrm{~d} E_\nu}\left(\tilde{E}_\nu, E_\nu\right) =  \sum_{j_1, j_2} \int_{t^-}^{t^+}\mathrm{d}t \int_{s^-}^{s^+}\frac{\mathrm{d}s}{2m_{j_1}} \frac{\mathrm{d} \sigma_{j j_1 \rightarrow i j_2}}{\mathrm{~d} t}\left(s, t\right) \ ,
\end{align}
where $n$ and $n^\prime$ are used to label the energy bins that will be integrated (summed over) in the {\tt nuSIprop} code. In the second step of each line, we used $s=2m_\nu\tilde{E}_{\nu}>0$, $t=-2m_\nu E_{\nu}<0$, where $\tilde{E}_{\nu}$ and $E_\nu$ are the incoming and outgoing energy of neutrino, respectively.

For the unparticle mediator model considered in this work, it is possible to complete most of the above integrals analytically. It is useful to define dimensionless integral variables, $S=s/\mu_{\rm IR}^2$ and $T=t/\mu_{\rm IR}^2$. We will present results for the $s$ and $t$ channels separately. The two contributions simply add to each other in $\Gamma$, $\tilde \alpha$ and $\alpha$.

\smallskip
\noindent\underline{\it $s$-channel contribution}

For $\Gamma_{i}^{n}$ we obtain
\begin{equation}
\Gamma_i^n    =  \frac{|\lambda|^4 |B_{d_u}|^2 |U_{\tau i}|^2 }{16 \pi} \bigg(\frac{\mu_{\rm IR}}{\Lambda}\bigg)^{4(d_u{-}1)} \sum_{j,k,l} \frac{|U_{\tau j}|^2|U_{\tau k}|^2 |U_{\tau l}|^2}{2m_j}  \int_{S^-}^{S^+} dS \Bigg[\frac{S}{\left(S {-} 1\right)^{2(2-d_u)} + S^2w^2} 
\Bigg],
\end{equation}
with $w= |\lambda|^2|B_{d_u}| /16\pi \big(\mu_{\rm IR} / \Lambda\big)^{2(d_u{-}1)}$. The integral can be expressed in terms of Gauss’ hypergeometric function
\begin{equation}
   \frac{1}{w^2} 
    \bigg[
     (S{-}1)\, {}_2F_1\left( 1, \frac{1}{4 {-} 2 d_u}, 1 {+} \frac{1}{4 {-} 2 d_u}, -\frac{ (S{-}1)^{2(2 - d_u)}}{w^2} \right)
     - \frac{(S{-}1)^2}{2} \, {}_2F_1\left( 1, \frac{1}{2 {-} d_u}, 1 {+} \frac{1}{2 {-} d_u}, -\frac{(S{-}1)^{2(2 {-} d_u)}}{w^2} \right)
    \bigg] \Bigg|^{S^+}_{S=S^-},
\end{equation}

For $\tilde{\alpha}$ we obtain
\begin{equation}
    \tilde{\alpha}_{ij}^n   =  \frac{|\lambda|^4 |B_{d_u}|^2 |U_{\tau i}|^2 |U_{\tau j}|^2}{16 \pi} \bigg(\frac{\mu_{\rm IR}}{\Lambda}\bigg)^{4(d_u{-}1)} \sum_{{j_1},{j_2}} \frac{|U_{\tau {j_1}}|^2 |U_{\tau {j_2}}|^2}{2m_{j_1}} \int_{T^-}^{T^+} dT\int_{-T}^{-T^+} dS \Bigg[\frac{2}{\left(S {-} 1\right)^{2(2-d_u)} + S^2w^2} 
\Bigg],
\end{equation}
the factor of 2 in the numerator of the integral is because two identical neutrinos of energy $E_\nu$ are generated through the s-channel. Here the integral can be completed as
\begin{align}
\frac{2}{w^2}\Bigg\{ 
&
-\left( T^+{-}T^- \right)\left(T^+{+}1\right) \, {}_2F_1\left( 1, \frac{1}{4 {-} 2 d_u}, 1 {+} \frac{1}{4 {-} 2 d_u}, -\frac{ (T^+{+}1)^{2(2 - d_u)}}{w^2} \right) + \\ 
& 
\Bigg[\frac{\left(T{+}1\right)^2}{2} \, {}_3F_2\left( \left\{1, \frac{1}{4{-}2 d_u}, \frac{1}{2 {-} d_u}\right\}, \left\{1{+}\frac{1}{4 {-} 2 d_u}, 1 {+} \frac{1}{2 {-} d_u}\right\}, -\frac{ (T{+}1)^{2(2 - d_u)}}{w^2} \right) \Bigg]\Bigg|^{T^+}_{T=T^-}
\Bigg\} \ .
\end{align}

For $\alpha$ we obtain
\begin{equation}
     \alpha_{ij}^{n,n^\prime}   =  \frac{|\lambda|^4 |B_{d_u}|^2 |U_{\tau i}|^2 |U_{\tau j}|^2}{16 \pi} \bigg(\frac{\mu_{\rm IR}}{\Lambda}\bigg)^{4(d_u{-}1)} \sum_{{j_1},{j_2}} \frac{|U_{\tau {j_1}}|^2 |U_{\tau {j_2}}|^2}{2m_{j_1}} \int_{T^-}^{T^+} dT\int_{S^-}^{S^+} dS \Bigg[\frac{2}{\left(S {-} 1\right)^{2(2-d_u)} + S^2w^2} 
\Bigg],
\end{equation}
where the integral is given by
\begin{align}
\frac{2}{w^2} \left( T^+{-}T^- \right) \Bigg[\left(S{-}1\right) \, {}_2F_1\left( 1, \frac{1}{4 {-} 2 d_u}, 1 {+} \frac{1}{4 {-} 2 d_u}, -\frac{ (S{-}1)^{2(2 - d_u)}}{w^2} \right)\Bigg] \Bigg|^{S^+}_{S=S^-}
\ .
\end{align}

\smallskip
\noindent\underline{\it $t$-channel contribution}

For $\Gamma_{i}^{n}$ we obtain
\begin{equation}
\Gamma_i^n = \frac{|\lambda|^4 |B_{d_u}|^2 |U_{\tau i}|^2 }{16 \pi} \bigg(\frac{\mu_{\rm IR}}{\Lambda}\bigg)^{4(d_u{-}1)} \sum_{j,k,l} \frac{|U_{\tau j}|^2|U_{\tau k}|^2 |U_{\tau l}|^2}{2m_j}  \int_{S^-}^{S^+} dS  \frac{-1 + \left(S + 1\right)^{2d_u{-}3} \big[(d_u{-}1)(2d_u{-}3)S^2 - (2 d_u{-}3) S + 1\big]}{(d_u{-}1)(2d_u{-}3)(2d_u{-}1)S^2 }  \ ,
\end{equation}
where the integral takes the analytic form
\begin{equation}
    \Bigg[       
 \frac{1 + \left(S + 1\right)^{2(d_u{-}1)} \big[(d_u{-}3/2)S - 1\big]}{(d_u{-}1)(2d_u{-}3)(2d_u{-}1)S} 
\Bigg]\Bigg|^{S^+}_{S=S^-} \ .
\end{equation}

For $\tilde{\alpha}$ we obtain
\begin{equation}
    \tilde{\alpha}_{ij}^n   =  \frac{|\lambda|^4 |B_{d_u}|^2 |U_{\tau i}|^2 |U_{\tau j}|^2}{16 \pi} \bigg(\frac{\mu_{\rm IR}}{\Lambda}\bigg)^{4(d_u{-}1)} \sum_{{j_1},{j_2}} \frac{|U_{\tau {j_1}}|^2 |U_{\tau {j_2}}|^2}{2m_{j_1}} \int_{T^-}^{T^+} dT\int_{-T}^{-T^+} dS \Bigg[\frac{T^2/S^2}{\left(-T {+} 1\right)^{2(2-d_u)}} + \frac{(S+T)^2/S^2}{\left(S{+}T {+} 1\right)^{2(2-d_u)}} 
\Bigg] \ ,
\end{equation}
Note the additional term is because, through $t$-channel scattering, the final state neutrino and antineutrino carry energies $E_\nu$ and $\tilde{E}_\nu-E_\nu$, respectively. In the charge-conjugation channel, their energies are swapped.
Both contribute to the neutrino flux.
CP is simply conserved for the tree-level processes evaluated here. The integral of the first term in the square bracket is
\begin{equation}
\small{ -\frac{
    \left( (T^-{-}1)^3(1 {-} T^+)^{2d_u}(2 {+} (2d_u{-}3)T^+) 
    + (1 {-} T^-)^{2d_u}(T^+{-}1)^2
    \left( 2 {+} 2(2d_u{-}3)T^-(1 {+} (d_u{-}1)T^-) 
    - (2d_u{-}1)(1 {+} (2d_u{-}3)T^-)T^+ \right) \right)
}{
    2(d_u{-}1)(2d_u{-}3)(2d_u{-}1)(T^-{-}1)^3(T^+{-}1)^2T^+
} \ ,}
\end{equation}
whereas the second term is not analytic and computed numerically.

Finally, for $\alpha$ we obtain
\begin{equation}
    \alpha_{ij}^{n,n^\prime}   =  \frac{|\lambda|^4 |B_{d_u}|^2 |U_{\tau i}|^2 |U_{\tau j}|^2}{16 \pi} \bigg(\frac{\mu_{\rm IR}}{\Lambda}\bigg)^{4(d_u{-}1)} \sum_{{j_1},{j_2}} \frac{|U_{\tau {j_1}}|^2 |U_{\tau {j_2}}|^2}{2m_{j_1}} \int_{T^-}^{T^+} dT\int_{S^-}^{S^+} dS \Bigg[\frac{T^2/S^2}{\left(-T {+} 1\right)^{2(2-d_u)}} + \frac{(S+T)^2/S^2}{\left(S{+}T {+} 1\right)^{2(2-d_u)}} 
\Bigg] \ ,
\end{equation}
where the first term of the integral can be done analytically
\begin{equation}
\footnotesize{\left(\frac{1}{S^+}  {-} \frac{1}{S^-}  \right)\frac{
\left(
-2(T^-{-}1)^{2d_u {-}3} \left( 1 {+} (2d_u{-}3)T^- \left( 1 {+} (d_u{-}1)T^- \right) \right)
+ 4d_u^2(T^+{-}1)^{2d_u{-}3}(T^+)^2
- 2(T^+{-}1)^{2d_u{-}3} \left(T^+ {-}1\left( 3 {-} 3T^+ {+} d_u( 5T^+{-}2) \right) \right)
\right)
}{
2(d_u{-}1)(2d_u{-}3)(2d_u{-}1)
}} \ ,
\end{equation}
whereas the second term is not analytic and computed numerically.

\subsection{C. 
IceCube constraint on unparticle-mediated neutrino self-interaction: posterior distributions}\label{appendix:triangleplot}

The model parameters with an unparticle mediator considered in this work are $\{\lambda, \Lambda, \mu_{\rm IR}, d_u\}$. We explore cases with $d_u=1.1, 1.2, 1.3$ and hold $\Lambda=100$\,GeV fixed. The corresponding UV cutoff scale for the unparticle spectral density function $\mu_{\rm UV}$ is calculated using Eq.~\eqref{eq:redundancy} and ranges between 700 GeV and 1 TeV. 
Another fundamental physics parameter is the sum over active neutrino masses $\sum m_{\nu}$, or the lightest neutrino mass $m_\nu^L$. 
We focus on the range $\sum m_{\nu}<0.12\,{\rm eV}$ to be consistent with cosmology and work with normal mass ordering.

The astrophysical parameters involved in our analysis are $\Phi_{\rm astro}$ and $\gamma_{\rm astro}$. They are used to define the UHE neutrino spectrum in the source term of the propagation equation, under the single power law assumption,
\begin{equation}
\frac{d\Phi_{6\nu}}{dE_\nu} = \Phi_{\rm astro} \left( \frac{E_\nu}{100\,\rm TeV} \right)^{-\gamma_{\rm astro}} \cdot 10^{-18}\,{\rm GeV}^{-1}{\rm cm}^{-2}{\rm s}^{-1}{\rm sr}^{-1} \ ,
\end{equation}
where $\Phi_{6\nu}$ is the sum of all neutrino and antineutrino fluxes.

\begin{figure}[h]
    \centering
    \includegraphics[width=0.8\linewidth]{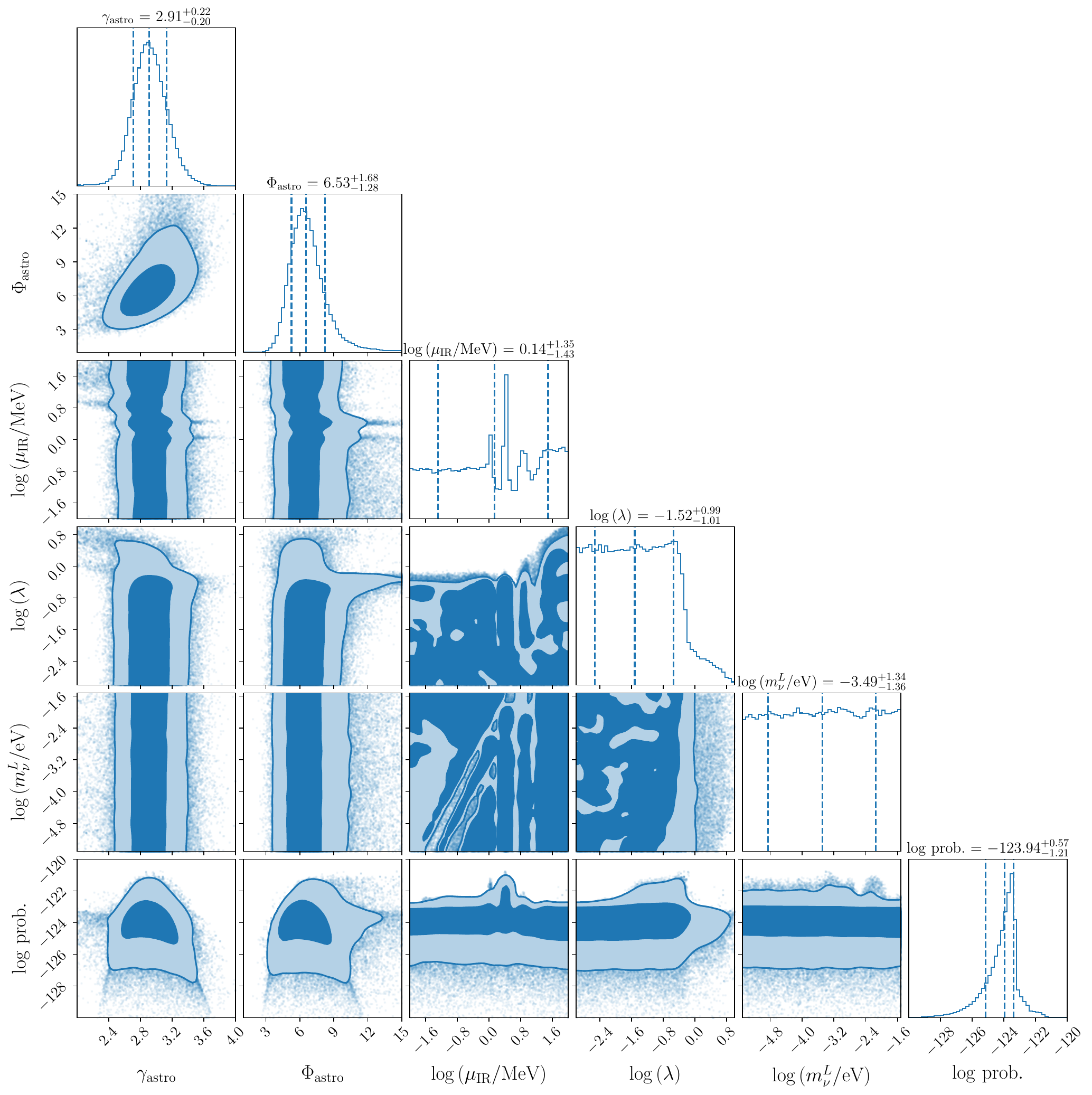}
    \caption{Posterior distributions of the parameters of interest for an unparticle scenario with $d_u = 1.2$. Diagonal panels project each parameter's one-dimensional marginalized distribution from the MCMC chain. The off-diagonal panels display the corresponding two-dimensional marginalized distributions, showing the correlations between parameters. Contours represent the $68\%$ and $95\%$ highest-posterior density regions.
    \label{fig:posterior_prob}}
\end{figure}

We scan over $\lambda$, $\mu_{\rm IR}$, $\sum m_{\nu}$, $\Phi_{\rm astro}$ and $\gamma_{\rm astro}$ and compare the predicted neutrino spectrum by the unparticle mediator model with the IceCube 7.5-year data. The corner plot in Fig.~\ref{fig:posterior_prob} shows the posterior distributions of the parameters for and unparticle case with $d_u=1.2$. The plots for $d_u=1.1$ (1.3) is similar except that the constraint on $\lambda$ is stronger (weaker).

\begin{figure}[h]
    \centering
    \includegraphics[width=0.45\linewidth]{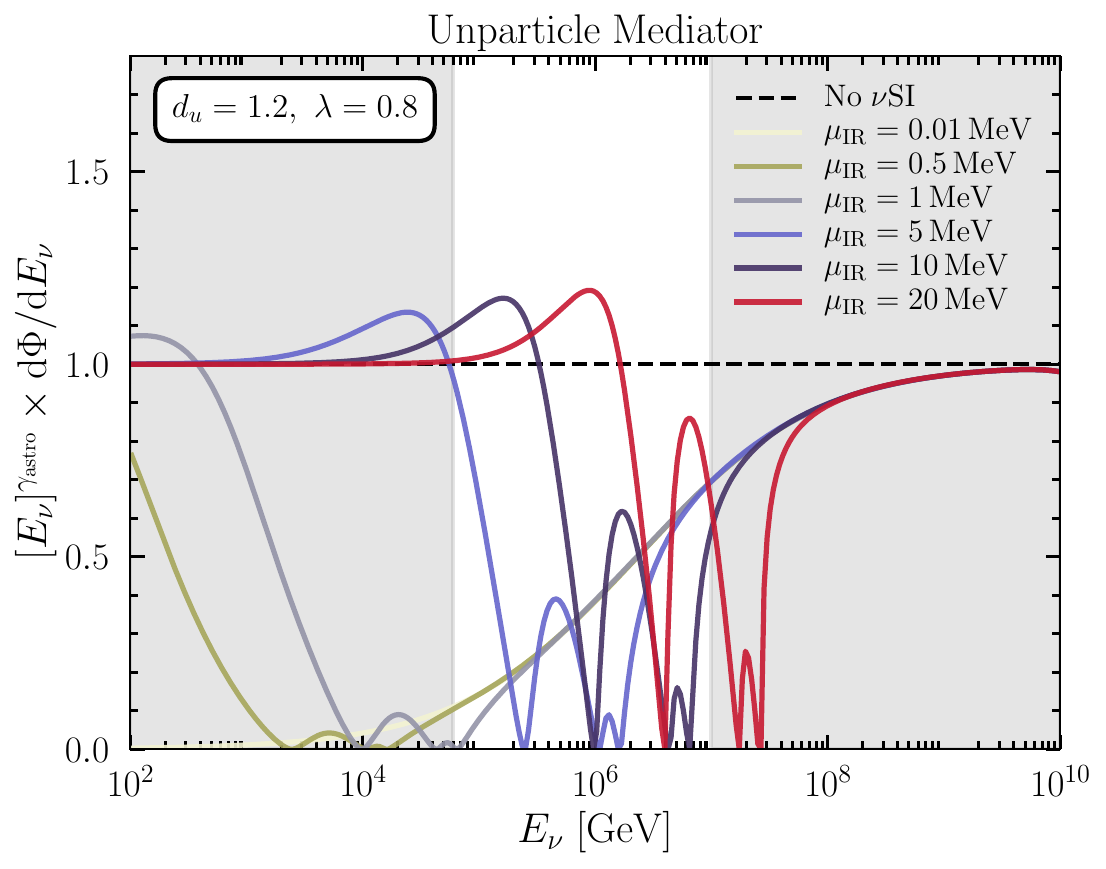}
    \includegraphics[width=0.45\linewidth]{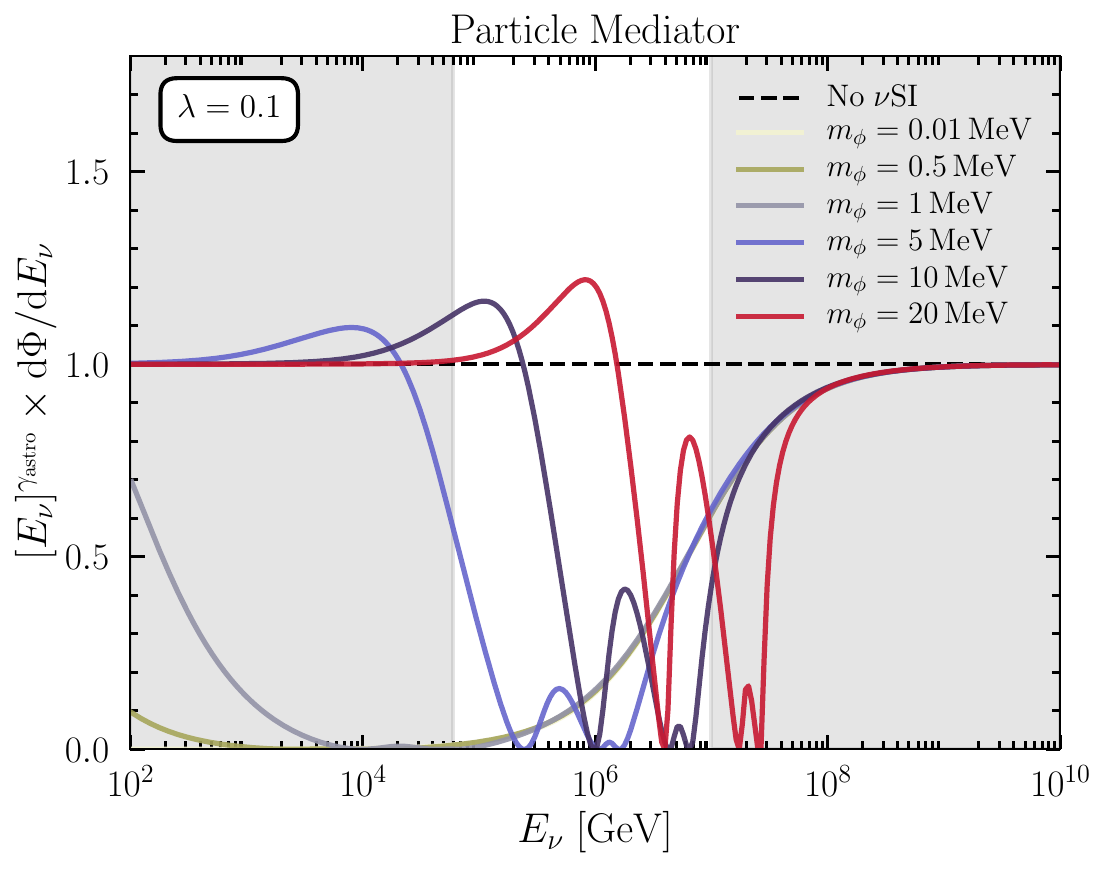}
    \caption{Neutrino flux at Earth in the presence of $\nu$SI for both (\textbf{left}) unparticle and (\textbf{right}) particle mediators. The left panel shows the absorption and regeneration of neutrinos as a function of incoming neutrino energy for different mass gaps of unparticle mediators with scaling dimension $d_u=1.2$ ($\lambda=0.8$, $\Lambda=100 \, \rm{GeV}$). The right panel shows a similar effect for particle mediators $\phi$ with various masses ($\lambda=0.1$). The gray-shaded regions are not considered in the HESE 7.5-year data analysis. Both plots assume a total neutrino mass $\sum m_{\nu} = 0.07 \, \rm{eV}$.
    \label{fig:nuSI_flux}}
\end{figure}

It is worth noting that in Fig.~\ref{fig:All_du} the IceCube limit does not go away for very small $\mu_{\rm IR}$ (or $m_\phi$). To address this point, we present Fig.~\ref{fig:nuSI_flux}, which shows the UHE neutrino flux that reaches the Earth. In the presence of neutrino self-interactions, the flux is obtained using {\tt nuSIprop}. 
The strongest absorption effects occur near $E_\nu \simeq \mu_{\rm IR}^2/(2m_\nu)$, where $m_\nu$ is one of the neutrino mass eigenvalues. For smaller $\mu_{\rm IR}$ (or $m_\phi$), the dips move toward the left and eventually out of the energy window ($600 \,\rm TeV$ -- $10 \, \rm PeV$) of cosmogenic neutrinos that IceCube makes an observation. 
On the other hand, we find that absorption effects away from the resonant dips, albeit less significant, are not negligible. 
For sufficiently small $\mu_{\rm IR}$ (or $m_\phi$) (see the golden and gray curves in Fig.~\ref{fig:nuSI_flux}), the neutrino flux in the above energy window asymptotes to a universal deficit behavior -- leading to a constraint on $|\lambda|$ that is insensitive to $\mu_{\rm IR}$ (or $m_\phi$).
This happens to both the particle and unparticle mediator cases.

\subsection{D. Invisible $Z$ width in presence of neutrino self-interaction -- particle and unparticle mediator cases}\label{appendix:ZDecay2}

In the case of a particle mediator for neutrino self-interaction, the low-energy Lagrangian takes the form
\begin{align}\label{eq:Lagrangian2}
    \mathcal{L}_{\rm eff.} = \frac{\lambda}{2} \bar{\nu}_\tau^c \mathbb{P}_L \nu_\tau \phi + \mathrm{h.c.} \ ,
\end{align}
where $\phi$ is a scalar particle, assumed to be real here for minimizing its contribution to $\Delta N_{\rm eff.}$. Again, we focus on the case of  $\nu_\tau$ self-interaction.
Eq.~\eqref{eq:Lagrangian2} is equivalent to Eq.~\eqref{eq:Lagrangian} in the limit where $d_u\to 1$.

The introduction of $\phi$-neutrino interaction in the above simplified model introduces extra radiative corrections to the electroweak Lagrangian in the lepton sector, which suffers from UV divergences. 
A recent work~\cite{Zhang:2024meg} pointed out that the UV divergences will disappear in gauge-invariant UV completions of Eq.~\eqref{eq:Lagrangian2} by virtue of the Ward identity. 
By assuming that all new particles (with a mass scale denoted by $M$) are heavy, except for $\phi$,
\cite{Zhang:2024meg} further used the decoupling argument to derive model-independent renormalized $Z\nu\bar\nu$ coupling at one-loop level. In the $M^2\gg |q^2|$ limit,
\begin{equation}
g_Z
= g_Z^0 - \frac{g_Z^0|\lambda|^2}{32\pi^2} \left\{ \ln\frac{M^2}{m_\phi^2} -\frac{3}{2} + 2\int_{0}^{1} dx \int_{0}^{1-x} dy \left[\ln \left(\frac{M^2}{(1-x-y)m_\phi^2-xy q^2}\right) + \frac{x y q^2}{(1-x-y)m_\phi^2-xy q^2} \right] \right\} \ ,
\end{equation}
where $q^\mu$ is the four-momentum carried by the $Z$ field, and $g_Z^0 = (\sqrt{2} G_F M_Z^2)^{1/2}$ is the tree-level $Z$ coupling.
Using this result, the $Z$-boson decay width via virtual $\phi$ exchange and real $\phi$ emission are given by
{\small \begin{align}
    &\delta \Gamma_{Z\to\nu_\tau\bar\nu_\tau} = - \frac{\sqrt{2} G_F M_Z^3 |\lambda|^2}{192\pi^3} \int_{0}^{1} dx \int_{0}^{1-x} dy \mathrm{Re}\left\lbrace \ln\frac{M^2}{m_\phi^2} + \ln\left[\frac{M^2}{(1-x-y)m_\phi^2 - xyM_Z^2}\right] + \frac{xyM_Z^2}{(1-x-y)m_\phi^2 - xyM_Z^2} - \frac{3}{2}  \right\rbrace \ ,\label{eq:app1}\\
    &\Gamma_{Z\to\nu_\tau\nu_\tau \phi} + \Gamma_{Z\to\bar\nu_\tau\bar\nu_\tau \phi^*} = \frac{\sqrt{2} G_F M_Z^3 |\lambda|^2}{192\pi^3} \left\lbrace \left[\frac{1}{2}\ln{\left(\frac{M_Z^2}{m_\phi^2}\right)} - \frac{11}{6}\right] + \frac{4m_\phi^2}{M_Z^2}\left[\ln{\left(\frac{M_Z^2}{m_\phi^2}\right)} - \frac{5}{8}\right] \right. \nonumber\\
&\hspace{3cm}\left. + \frac{m_\phi^4}{M_Z^4} \left[\ln{\left(\frac{M_Z^2}{m_\phi^2}\right)} \left(
\ln{\left(\frac{M_Z m_\phi}{M_Z^2+m_\phi^2}\right)} + \frac{3}{2}\right) + \frac{9}{2}
+{\rm Li}_2\left( \frac{M_Z^2}{M_Z^2+m_\phi^2}\right)
-{\rm Li}_2\left( \frac{m_\phi^2}{M_Z^2+m_\phi^2}
\right)
\right] - \frac{m_\phi^6}{6M_Z^6}\right\rbrace \ .\label{eq:app2}
\end{align}}
We have neglected neutrino masses throughout the calculation. In the $m_\phi\to0$ limit, the infrared-divergent terms (proportional to $\ln m_\phi$) cancel in the sum of Eqs.~(\ref{eq:app1}, \ref{eq:app2}) as found in~\cite{Brdar:2020nbj, Dev:2024twk}.
However, we note that for the three-body decay considered here, the two final state (anti-)neutrinos are identical particles, thus the decay matrix element contains two terms that interfere with each other, originating from two ways of assigning the neutrino momenta. If the $\phi$ coupling were flavor changing and the two final state (anti-)neutrinos had different flavors, we reproduce the same result as those in Refs.~\cite{Brdar:2020nbj,Dev:2024twk}.

\begin{figure}[h]
    \centering
    \includegraphics[width=0.45\linewidth]{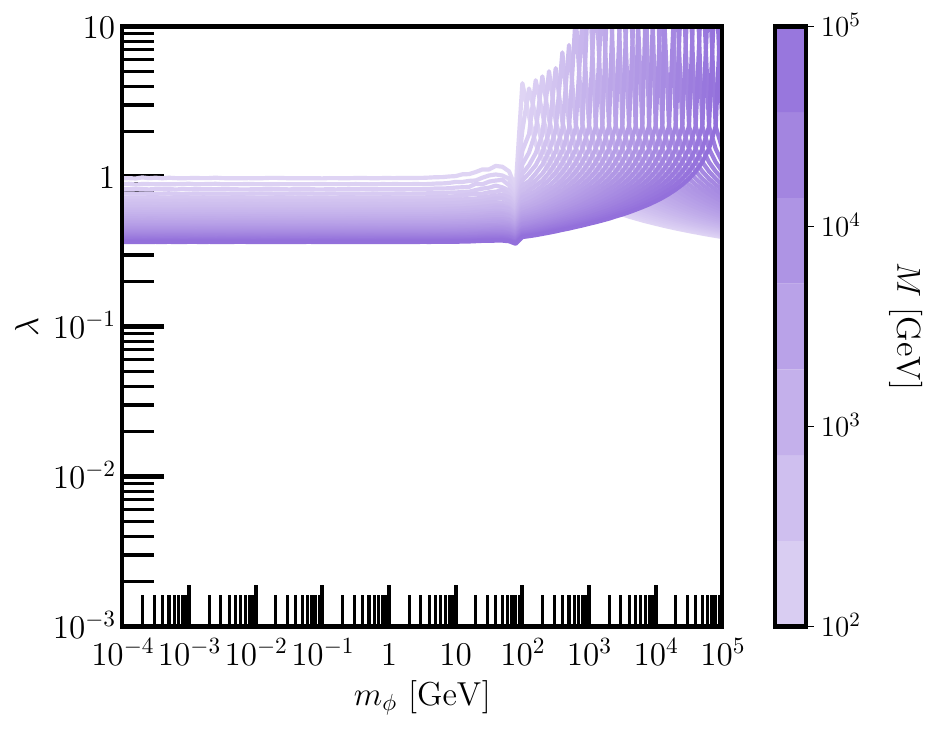}
    \caption{Constraints on the particle case as a function of $m_\phi$ and $\lambda$ arising from measurements of the invisible width of the $Z$ boson. Here we allow $M$, the mass scale of UV physics, to vary between $100$~and~$10^5$~GeV -- lighter (darker) lines correspond to smaller (larger) values of $M$.
    \label{fig:ZwidthUV}}
\end{figure}

The invisible $Z$-decay rate has a logarithmic dependence on the UV mass scale $M$, which affects the corresponding constraint on $\lambda$. Fig.~\ref{fig:ZwidthUV} shows the upper bound on $\lambda$ for several choices of $M$. For small $m_\phi \ll M_Z, M$, the upper bound is insensitive of $m_\phi$ (as a result of the cancellation of $\ln m_\phi$ terms) and gets stronger for larger $M$. The upper bound weakens substantially for $m_\phi\sim M$.

Next, we generalize the above $Z$-boson decay rate to the case of a neutrinophilic unparticle mediator introduced in~\cref{eq:Lagrangian}.
Following the discussions in the main text, the difference between a particle and unparticle can be accounted for by a nontrivial Källén–Lehmann spectral density function rather than a Dirac $\delta$-function,
\begin{align}\label{eq:replacement}
\begin{array}{ccc}
{\rm particle} & & {\rm unparticle} \\
(2\pi)\delta(m^2-m_\phi^2)  & \hspace{0.5cm}\to & \quad \rho_{\rm KL}(m^2) \ ,
\end{array}
\end{align}
where $\rho_{\rm KL}(m^2)$ is given by Eq.~\eqref{eq:KL}. The generalization of the $Z$-boson invisible decay width to the model with unparticle-neutrino interaction (see Eq.~\eqref{eq:Lagrangian}) goes as follows. 
For the Feynman propagator in momentum space, Eq.~\eqref{eq:replacement} implies the replacement
\begin{align}
\frac{i}{p^2 - m_\phi^2 + i\varepsilon} = \int \frac{dm^2}{2\pi} (2\pi) \delta(m^2-m_\phi^2) \frac{i}{p^2 - m^2 + i\varepsilon} \quad \to \quad \int_{\mu_{\rm IR}^2}^{\mu_{\rm UV}^2} \frac{dm^2}{2\pi} \rho_{\rm KL}(m^2) \frac{i}{p^2 - m^2 + i\varepsilon} \ .
\end{align}
For the $Z\to\nu\bar\nu$ decay, it is useful to note that the one-loop diagrams interfere with the tree-level one, and
the loop diagram contains only one $\phi$ (or $\mathcal{U}$) propagator. Therefore, we simply need to plug the $\delta \Gamma_{Z\to\nu\bar\nu}$ decay rate Eq.~\eqref{eq:app1} into the following integral
\begin{align}\label{appeq:int1}
\frac{1}{\Lambda^{2d_u-2}}\int_{\mu_{\rm IR}^2}^{\mu_{\rm UV}^2} \frac{dm^2}{2\pi} \rho_{\rm KL}(m^2) \ ,
\end{align}
after rewriting $m_\phi^2\to m^2$.
The prefactor arises from the dimensionful parameter in the unparticle interaction Lagrangian Eq.~\eqref{eq:Lagrangian}. This results in Eq.~\eqref{eq:7} in the main text.

For the process with real $\mathcal{U}$ emission, we need to implement a similar replacement for the final state phase space integral. 
In the particle case, the integral can be written as
\begin{align}
\int\frac{d^3 \vec{p}}{(2\pi)^3} \frac{1}{2 \sqrt{|\vec{p}|^2+m_\phi^2}} &= \int\frac{d^3 \vec{p}}{(2\pi)^3} \int \frac{dp^0}{2\pi} (2\pi) \delta\left((p^0)^2 - |\vec{p}|^2 - m_\phi^2\rule{0mm}{4mm}\right) \Theta(p^0) \nonumber \\
&=\int\frac{d^3 \vec{p}}{(2\pi)^3}\int \frac{dp^2}{2\pi} \frac{1}{2p^0}  (2\pi) \delta\left(p^2 - m_\phi^2\right) \Theta(p^0) \nonumber \\
&=\int\frac{dm^2}{2\pi}\int\frac{d^3 \vec{p}}{(2\pi)^3}  \frac{1}{2 \sqrt{|\vec{p}|^2+m^2}}  (2\pi) \delta\left(m^2 - m_\phi^2\right) \Theta(p^0) \ ,
\end{align}
where in the second step, we use the identity $dp^2 = 2p^0dp^0$, which is derived by differentiating $p^2 \equiv (p^0)^2 - |\vec{p}|^2$ and holding $\vec{p}$ fixed.
In the last step, we rename $p^2\to m^2$ and interchanged the order of two integrals.
Next, following the rule of Eq.~\eqref{eq:replacement}, the final state phase space integral for unparticle must take the form
\begin{align}
\int\frac{dm^2}{2\pi} \rho_{\rm KL}(m^2) \int\frac{d^3 \vec{p}}{(2\pi)^3}  \frac{1}{2 \sqrt{|\vec{p}|^2+m^2}}  \ .
\end{align}
Therefore, for the $Z\to \nu\nu\mathcal{U}^*$ and $Z\to \bar\nu\bar\nu\mathcal{U}$ decays, we insert Eq.~\eqref{eq:app2} into the integral
\begin{align}\label{appeq:int2}
\frac{1}{\Lambda^{2d_u-2}}\int_{\mu_{\rm IR}^2}^{\mu_{\rm UV}^2} \frac{dm^2}{2\pi} \rho_{\rm KL}(m^2) \ ,
\end{align}
after rewriting $m_\phi^2\to m^2$.
This gives Eq.~\eqref{eq:8} in the main text.

Because the two spectral integrals Eqs.~\eqref{appeq:int1} and \eqref{appeq:int2}, are identical, in the $m_\phi\to0$ limit, the cancellation of infrared divergences between virtual $\phi$ exchange and real $\phi$ emission in the $Z$-boson decay still holds for the unparticle case.

\end{document}